\documentclass[aps,prl,twocolumn,superscriptaddress,showpacs]{revtex4}
\usepackage{amsmath}
\usepackage{amssymb}
\usepackage{graphicx}
\usepackage{txfonts}
\usepackage{color}

\begin{document}
\title{First-principles investigation of transient current of molecular devices\\ by using complex absorbing potential}

\author{Lei Zhang, Jian Chen and Jian Wang$^*$}
\address{Department of Physics and the Center of Theoretical and
Computational Physics, The University of Hong Kong, Hong Kong, China}

\begin{abstract}
Based on the non-equilibrium Green's function (NEGF) coupled with density function theory (DFT), namely, NEGF-DFT quantum transport theory, we propose an efficient formalism to calculate the transient current of molecular devices under a step-like pulse from first principles. By combining NEGF-DFT with the complex absorbing potential (CAP), the computational complexity of our formalism (NEGF-DFT-CAP) is proportional to $\emph{O}(N)$ where $N$ is the number of time steps in the time-dependent transient calculation. Compared with state-of-the-art algorithm of first principles time-dependent calculation that scales with at least $N^2$, this order N technique drastically reduces the computational burden making it possible to tackle realistic molecular devices. To ensure the accuracy of our method, we carry out the benchmark calculation compared with exact NEGF-TDDFT formalism and they agree well with each other. As an illustration, we investigate the transient current of molecular device Al-C$_3$-Al from first principles.
\end{abstract}

\pacs{
71.15.Mb,  
72.10.-d,  
85.65.+h,   
73.63.-b   
}

\maketitle

\section{Introduction}

With the advance of nanofabrication techniques, people can fabricate the nano-devices using single atoms or
molecules from bottom-up approach, which leads to a new field of molecular
electronics\cite{Aviram,Chen,Collier,Joachim,Heath,Petta,Tao}. Many experiments have been performed to measure quantum transport properties of molecular devices\cite{Chen,Collier,Joachim,Heath,Petta,Tao}. At the same time, people have made a lot of research efforts to understand these properties from first principles\cite{Ventra,Taylor,Brandbyge,xue}. At present stage, quantitative agreement between theoretical first principles calculations and experiment results can be reached when the system is in the steady state regime under external DC bias\cite{Kaun1,Kaun2,Frederiksen,Lee}. Besides the DC steady state problem, the question of how fast a molecular device can turn on and off is also an important issue, which attracts a lot research attention recently\cite{Jauho1,Gross1,Stefanucci,lei,Maciejko,Zheng}. This kind of question can be answered by studying the dynamic response of molecular devices by sending a step like pulse from the electrodes. For this problem, exact solution of transient current was obtained by Wingreen et al\cite{Jauho1} in the wide-band limit using non-equilibrium Green's function (NEGF). Recently this solution has been extended to the regime of finite band width of electrodes\cite{Maciejko}. When applying this exact NEGF solution to molecular devices for calculation of transient current as a function of time, there is a huge computational cost that scales with $O(N^3)$ due to the triple integral over energy, where $N$ is the number of time steps. In addition, there are many quasi-poles near the energy axis making the integration very difficult to converge. On the other hand, the theoretical prediction of the transient dynamics of molecular devices from first principles can be addressed by numerically solving scattering wave function or non-equilibrium Green's function (NEGF) combined with time dependent density functional theory (TDDFT)\cite{Gross1,Stefanucci,lei}. These methods again are very time-consuming for transient current calculation although the scaling has been reduced to $N^2 (log_2 N)^2$. Therefore, to speed up calculation, various approximate schemes were proposed to calculate time dependent transient current of molecular devices such as wide band approximation\cite{Zheng}. Another approximate scheme based on exact NEGF solution was also proposed and applied to calculate transient current of molecular devices which is very efficient and goes beyond the wideband limit\cite{Bin1}. Despite of these efforts, time-dependent calculation of transient current for molecular devices is still a challenge on the computational resources. Due to the importance of molecular electronics, it is timely to overcome this problem so that realistic transient dynamics calculations can be performed on molecular devices from first principles. In this paper, we propose a linear scaling $\emph{O}(N)$ scheme to calculate the time dependent transient current by combining complex absorbing potential (CAP) method with exact solution based on NEGF \cite{Maciejko} and DFT theory (NEGF-DFT-CAP).

The CAP was initially used to simulate the time-dependent evolution of wave function of finite systems in one and two dimensions\cite{Kosloff}. Recently, CAP was employed to study the transport problem of molecular device from first principles\cite{Henderson,Zhang,Varga1,Varga2} using a transmission free CAP\cite{Manolopoulos}. By adding an energy independent CAP in lead regions, the transport problem in a infinite open system can be reduced to that of a finite simulation region. Moreover, one can obtain an effective wideband-like formula to calculate dc transport quantities such as the transmission coefficient. We note that in general ac transport properties do not assume the wideband form (effective self-energy does not depend on the energy) in the presence of CAP. Fortunately, for the step-like pulse, we are able to cast the exact NEGF solution for transient current into a wideband form using CAP which enables us to speed up the calculation tremendously. In fact,
when CAP is implemented into the exact NEGF solution the amount of calculation scales like $c N n^3$ where $N$ is the number of time steps, $n$ is the dimension of the Hamiltonian in the whole simulation region including the CAP region, and $c$ is a constant of order of a few hundred. With this order $N$ method at hand, the first principles calculation of transient current of realistic molecular devices are within the reach. We have applied our formalism to molecular devices and carried out the benchmark calculation for a one-dimensional atomic chain which agrees with the result from exact numerical calculation. Furthermore, we have investigated transient dynamics of a 3D molecular device and calculated transient current at two different bias voltages. It was found that the transient current involves many time scales showing that the wideband limit is a bad approximation for molecular devices.

The paper is organized as follows. In section $2$, we will first introduce the formalism of complex absorbing potential (CAP) and briefly discuss its application in DC transport calculations. Then we will discuss how to apply the CAP to calculate the time dependent transient current of molecular device under upward step-like pulse. In section $3$, benchmark comparisons with NEGF-TDDFT method is presented. Then numerical calculation of transient current of the Al-C$_3$-Al molecular device is given. Finally, section $4$ serves as discussion and conclusion part.

\section{Theoretical formalism}

\subsection{Complex absorbing potential (CAP)}

As shown in figure $1$, a typical two terminal device consists of the central scattering region connected by two semi-infinite external leads along the transport $z$ direction. The corresponding Hamiltonian of the whole system can be expressed as a tri-diagonal block matrix
\begin{equation}
H=\left[\begin{array}{ccc}
        H_{LL} & H_{LC} & 0\\
        H_{CL} & H_{CC} & H_{CR}\\
        0 & H_{RC} & H_{RR}\\
\end{array}\right],\label{eqH}
\end{equation}
where $H_{\alpha\alpha},\alpha=L,R$ is the semi-infinite Hamiltonian of lead $\alpha$. In order to study the transport properties of this open system, one is actually solving the scattering problem of the infinite dimension. In the framework of non-equilibrium Green's function (NEGF), one calculates various Green's function of the central region and the effect of the leads is taken into account by the self-energy. For instance, the retarded Green's function of central region in energy domain is defined as
\begin{equation}
G^r_{CC}(E)=(E-H_{CC}-\sum_{\alpha=L,R}\Sigma^r_{\alpha}(E))^{-1},\label{tradG}
\end{equation}
where $\Sigma^r_{\alpha}(E)$ is the self energy of lead $\alpha$
\begin{equation}
\Sigma^r_{\alpha}(E)=H_{C\alpha}g^r_{\alpha\alpha}(E)H_{\alpha C},\label{self1}
\end{equation}
and $g^r_{\alpha\alpha}(E)$ is the retarded Green's function of the corresponding lead $\alpha$
\begin{equation}
g^r_{\alpha\alpha}(E)=(E-H_{\alpha \alpha}+i0^+)^{-1}.
\end{equation}
After obtaining the retarded Green's function, one can calculate various transport quantities, such as transmission coefficient
\begin{equation}
T(E) = {\rm Tr}[\Gamma_LG^r_{CC}\Gamma_RG^a_{CC}].
\end{equation}
Here $\Gamma_{\alpha}(E)=i(\Sigma^r_{\alpha}(E)-\Sigma^a_{\alpha}(E))$ is the linewidth function of lead $\alpha$. In the numerical calculation, the energy dependent self-energy can be calculated by using the iterative or quadratic eigenvalue approaches\cite{Sancho,Sanvito}. To distinguish from the CAP method, we will refer the above method as exact method.

\begin{figure}
\includegraphics[width=8.5cm,height=2.1cm]{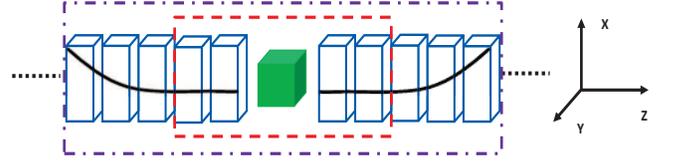}
\caption{Schematic plot of a two terminal molecular device. The
device consists of central molecular part (green solid cube) and two semi-infinite leads, which will extend to the $\pm \infty$. The black solid lines represent the complex absorbing potential added to both lead regions. The region enclosed by red dashed line is the central region; the purple dashed dot line encloses the central region plus complex absorbing potential region in the leads.} \label{fig1}
\end{figure}
The idea of CAP method is to replace the infinite system by a finite dimension using a transparent boundary condition that absorbs the incident wave function completely. In the application of CAP to the quantum transport problem, CAP is added to a finite lead region (called CAP region) outside of central scattering region. Usually, the effectiveness of CAP on absorbing the incident wave depends on the length of the CAP region. The reduction of reflection can be improved by increasing the length of CAP region in a controlled way. Note that the advantage of CAP method over the exact method relies on the fact that the CAP does not depends on energy while the self-energy of the exact method does. Using this property, the poles of Green's function for CAP method can be obtained easily. Therefore the convergence problem of energy integral in calculating transient current is solved. In the numerical calculation, we adopt a type of optimized transmission-free CAP form given in Ref. \cite{Manolopoulos},
\begin{equation}
W(z)=\frac{\hbar^2}{2m}(\frac{2\pi}{\Delta z})^2f(z),
\end{equation}
where $f(z)$ is defined as
\begin{equation}
f(z)=\frac{4}{c^2}((\frac{\Delta z}{z_2-2z_1+z})^2+(\frac{\Delta z}{z_2-z})^2-2),
\end{equation}
and $\Delta z=z_2-z_1$ is the range of CAP along transport $z$ direction, $z_1$ and $z_2$ are the starting and ending
points of CAP region at each lead, respectively. Here $c$ is a constant taken to be $2.62$, $m$ is the electron's mass. As shown in figure \ref{fig1}, the CAP region starts from several buffer layers away from the central molecular region. Going deep into the lead, the strength of absorbing potential increases and $f(z)\rightarrow \infty$ when $z$ approaches the end point $z_2$. This truncates the semi-infinite lead into a finite one. Therefore, the numerical simulation region becomes finite, i.e., the region enclosed by the purple dashed dot line in figure $1$. In the first principles calculation, LCAO basis set is usually adopted. Then one has to calculate the matrix element of CAP in orbital space
\begin{equation}
W_{\mu\nu}=\int \phi^*_{\mu}(x,y,z)W(z)\phi_{\nu}(x,y,z) dx dy dz,
\end{equation}
where $\phi_{\mu}$ is the atomic orbital.

Within the CAP method (all quantities are labeled with prime), the retarded Green's function of lead $\alpha$ can be defined as\cite{Varga1}
\begin{equation}
g^{r'}_{\alpha \alpha}(E) = (E-H_{\alpha\alpha}+iW_{\alpha})^{-1}.\label{EqLeadGr}
\end{equation}
Since the lead is effectively truncated, $H'_{\alpha\alpha}=H_{\alpha\alpha}-iW_{\alpha}$ is a matrix of finite dimension shown in figure $1$. The retarded Green's function of whole system including the CAP region can be expressed as
\begin{equation}
G^{r'} = (E - H + i\sum_{\alpha}W_{\alpha}')^{-1}\label{eqG0}
\end{equation}
with $W'_{L}=\left[\begin{array}{ccc}
        W_{L} & 0 & 0\\
        0 & 0 & 0\\
        0 & 0 & 0
\end{array}\right]$ and $W'_{R}=\left[\begin{array}{ccc}  0 & 0 & 0\\ 0 & 0 & 0\\0 & 0 & W_{R} \end{array}\right]$.

Although the lead region (CAP region) is finite, we can still use the concept of self-energy and obtain an effective retarded Green's function of the central region within CAP method
\begin{equation}
G^{r'}_{CC} = (E - H_{CC}-\sum_{\alpha}\Sigma^{r'}_\alpha)^{-1},\label{eqG1}
\end{equation}
where self energy $\Sigma^{r'}_\alpha$ is given by\cite{datta}
\begin{equation}
\Sigma^{r'}_{\alpha}(E)=H_{C\alpha}g^{r'}_{\alpha\alpha}(E)H_{\alpha C}.\label{trans1}
\end{equation}
It is easy to show that the linewidth function is written as\cite{Varga3}
\begin{equation}
\Gamma'_{\alpha} = 2H_{C\alpha}g^{r'}_{\alpha \alpha}W_{\alpha}g^{a'}_{\alpha\alpha}H_{\alpha C}=
2H_{C\alpha}g^{a'}_{\alpha \alpha}W_{\alpha}g^{r'}_{\alpha\alpha}H_{\alpha C}.\label{Eqgama}
\end{equation}
Since the self energy calculated by CAP method is the same as that obtained from the traditional method\cite{Varga1}, various Green's functions in the central (physical) region shown in figure \ref{fig1} should also be the same as that given in Eq. (\ref{tradG}). At this level, the self-energy of the Green's function of the central scattering region $G^{r'}_{CC}$ depends on energy. In the following, we give a simple derivation on transmission coefficient using CAP method in an effective wide band limit (WBL) form. Starting from the traditional definition of transmission coefficient of Eq. (\ref{trans1}) together with Eq. (\ref{Eqgama})
\begin{eqnarray}
T(E)&=& {\rm Tr}[\Gamma'_LG^{r'}_{CC}\Gamma'_RG^{a'}_{CC}]\nonumber\\
&=& 4{\rm Tr}[H_{CL}g^{a'}_{LL}W_{L}g^{r'}_{LL}H_{LC}G^{r'}_{CC} H_{CR}g^{r'}_{RR}W_{R}g^{a'}_{RR}H_{RC}G^{a'}_{CC}]\nonumber\\
& =& 4{\rm Tr}[W_{L}{G}^{r'}_{LR}W_{R}{G}^{a'}_{RL}]  = 4{\rm Tr}[W'_{L}{G}^{r'}W'_{R}{G}^{a'}]\label{EqT1}
\end{eqnarray}
where we have defined the following Green's function of the whole system including the CAP region (see the Appendix A for derivation)
\begin{eqnarray}
{G}^{r'}_{LR}=g^{r'}_{LL}H_{LC}G^{r'}_{CC}H_{CR}g^{r'}_{RR}.
\end{eqnarray}
In order to calculate transmission coefficient of Eq. (\ref{EqT1}), one only needs to know $G^{r'}$ which is defined in the whole system including the CAP region with $W'_{\alpha}$ an effective energy independent self energy. Note that this effective WBL form is only valid in DC case. In the case of AC transport, one may not have similar WBL form and one has to deal with it case by case.

In terms of the lesser Green's function, one can calculate the charge density in the central region. We also start from the traditional definition of the lesser Green's function
\begin{eqnarray}
G^{<}_{CC}(E)&&= i\sum_{\alpha}f_{\alpha}G^{r'}_{CC}\Gamma'_{\alpha}G^{a'}_{CC}\nonumber\\
&&= 2i\sum_{\alpha}f_{\alpha}G^{r'}_{C\alpha}W_{\alpha}G^{a'}_{\alpha C}\nonumber\\
&&= 2i\sum_{\alpha}f_{\alpha}[{G}^{r'}{W}'_{\alpha}{G}^{a'}]_{CC},
\label{gcc}
\end{eqnarray}
where we have used Eq. (\ref{Eqgama}) and $G^{r'}_{C\alpha}=G^{r'}_{CC}H_{C\alpha}g^{r'}_{\alpha\alpha}$ (see Eq.(3.5.13) in Ref.\cite{datta}); $f_\alpha$ is the Fermi distribution function of lead $\alpha$.

\subsection{Time dependent transient current with upward step like pulse}

The exact solution of time dependent current for step like pulse based on NEGF has been given by Maciejko et al \cite{Maciejko}. This formalism can be combined with DFT to calculate transient current in molecular devices\cite{Bin1,Xing1}. In the following, we will combine the exact solution with DFT and CAP to obtain an order $O(N)$ scheme (NEGF-DFT-CAP) for calculating time dependent current under upward step like pulse. Downward step and square like pulses can also be treated in a similar fashion.

To begin with, we will derive an equivalent time dependent current formula. Starting from the equation of motion for lesser Green's function\cite{Jauhobook}, we have
\begin{eqnarray}
i\frac{\partial}{\partial t}{G}_{CC}^<(t,t')&=&{H}_{CC}(t){G}_{CC}^<(t,t')+\int^{t}_{0}[{ \Sigma}^<(t,t_1){G}_{CC}^a(t_1,t')\nonumber\\
&&+{{\Sigma}^r(t,t_1)G}^<(t_1,t')]dt_1,\label{lGreen1}
\end{eqnarray}
and
\begin{eqnarray}
-i\frac{\partial}{\partial t'}{G}_{CC}^<(t,t')&=&{G}_{CC}^<(t,t'){H}_{CC}(t')+\int^{t}_{0}[{G}_{CC}^r(t,t_1){ \Sigma}^<(t_1,t')\nonumber\\
&&+G^<(t,t_1){\Sigma}^a(t_1,t')]dt_1.\label{lGreen2}
\end{eqnarray}
Then subtracting equation (\ref{lGreen2}) by equation (\ref{lGreen1}) and setting $t'=t$, we can arrive at
\begin{eqnarray}
I_{op}(t)={H}_{CC}(t){G}_{CC}^<(t,t)-{G}_{CC}^<(t,t){H}_{CC}(t)-i\frac{\partial}{\partial t}{G}_{CC}^<(t,t),\label{current1}
\end{eqnarray}
where we have defined
\begin{eqnarray}
I_{op}(t)&\equiv&\int^{t}_{0}[{G}_{CC}^r(t,t_1){\Sigma}^<(t_1,t)+G^<(t,t_1){\Sigma}^a(t_1,t)
\nonumber\\
&&-{ \Sigma}^<(t,t_1){G}_{CC}^a(t_1,t)-{{\Sigma}^r(t,t_1)G}^<(t_1,t)]dt_1.\label{current2}
\end{eqnarray}
which is a matrix. Note that the terminal current $I_{\alpha}(t)$\cite{Jauho} can be obtained from $I_{op}(t)$. To do that, two auxiliary projection matrices are introduced
\begin{eqnarray}
{\bar {\Gamma}}_{L}=
\left[\begin{array}{ccc}
        {1}_{L} & 0 & 0\\
        0 & 0 & 0\\
        0 & 0 & 0
\end{array}\right],
\qquad
{\bar {\Gamma}_{R}}=
\left[\begin{array}{ccc}
        0 & 0 & 0\\
        0 & 0 & 0\\
        0 & 0 & {1}_{R}
\end{array}\right],
\end{eqnarray}
where $\alpha$ denote outermost unit cell layer in the buffer layers of central region and ${1}_{L/R}$ is the unit matrix with dimension equal to the size of unit cell of left and right lead respectively so that $\Sigma^r_L = {\bar {\Gamma}}_{L} \Sigma^r {\bar {\Gamma}}_{L}$. Finally we have
\begin{eqnarray}
I_{\alpha}(t)={\rm Tr}[{\bar{\Gamma}}_{\alpha}I_{op}(t){\bar{\Gamma}}_{\alpha}].\label{current3}
\end{eqnarray}
From (\ref{current1}), we can use the following formula to calculate the time-dependent terminal current $I_{\alpha}(t)$,
\begin{equation}
I_{\alpha}(t)=2\mathrm{Re}\mathrm{Tr}[
{\bar{\Gamma}}_{\alpha}{H}_{CC}(t){G}^<_{CC}(t,t){\bar{\Gamma}}_{\alpha}]-i\mathrm{Tr}[
{\bar{\Gamma}}_{\alpha}\partial_t{G}_{CC}^<(t,t){\bar{\Gamma}}_{\alpha}],\label{tdc}
\end{equation}
where $G^<_{CC}(t,t)$ is the time dependent lesser Green's function of central region with equal time. To calculate $I_{\alpha}(t)$, one has to know the time dependent Hamiltonian $H(t)$ and calculate time dependent lesser Green's function $G^<_{CC}(t,t)$. Since the external bias is the upward step like pulse in our problem, then the time dependent Hamiltonian can be obtained as follows. When time $t<0$, $H(t<0)=H_{eq}$ is equilibrium Hamiltonian without bias and $H(t\geq0)=\theta(t)H_{neq}$ that $H_{neq}$ is the self consistent non-equilibrium Hamiltonian under DC bias. As for the time dependent lesser Green's function within CAP method, it can written as
\begin{equation}
{G}^<_{CC}(t,t)=2i\sum_{\alpha}\int\frac{d\omega}{2\pi}
f(\omega)[{A}^{'}_{\alpha}(\omega,t)W^{'}_{\alpha}{A}^{'\dagger}_{\alpha}(\omega,t)]_{CC},\label{Gl}
\end{equation}
where we have used the spectral function ${A}^{'}_{\alpha}(\omega,t)$\cite{Jauho}
\begin{equation}
{A}^{'}_{\alpha}(\epsilon,t)\equiv\int^t_{-\infty}dt'e^{i \epsilon (t-t')}e^{i \int^t_{t'} dt_1\Delta_{\alpha}(t_1)}{G}^{r'}(t,t'),
\end{equation}
where $\Delta_{\alpha}(t)$ is the time dependent external bias. Note that ${A}^{'}_{\alpha}(\epsilon,t)$ has the same dimension with ${G}^{r'}$ that is defined in CAP method. Then the key issue is how to calculate quantity ${A}^{'}_{\alpha}(\epsilon,t)$ efficiently. From the analytic expression of ${A}_{\alpha}(\epsilon,t)$ given in Ref. \cite{Maciejko}, we can derive the spectral function in the CAP form (see Appendix for derivation),
\begin{equation}
\begin{split}
{A}^{'}_{\alpha}(\epsilon,t)&=\bar{{G}}^{r'}(\epsilon+\Delta_{\alpha})
-\int\frac{d\omega}{2\pi i}\frac{e^{-i(\omega-\epsilon)t}\bar{{G}}^{r'}(\omega+\Delta_{\alpha})}{\omega-\epsilon+\Delta_{\alpha}-i0^+}\\&
[\frac{\Delta_{\alpha}}{\omega-\epsilon-i0^+}+{\Delta}\tilde{{G}}^{r'}(\epsilon)],\label{EqA}
\end{split}
\end{equation}
\begin{figure}[!ht]
\includegraphics[width=8.5cm,height=1.8cm]{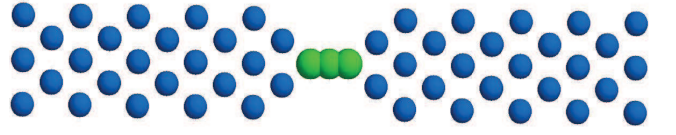}
\caption{Schematic diagram of a molecular device Al-C$_3$-Al. The
device consists of three carbon atoms chain coupled to the perfect
aluminium atomic electrodes which will extend to the reservoirs at
$\pm \infty$, where the current is collected.} \label{fig2}
\end{figure}
where ${\Delta}=H_{neq}-H_{eq}$ is internal potential change due to the external bias. Equilibrium and non-equilibrium retarded Green's function are defined as
\begin{equation}
\tilde{{G}}^{r'}(\epsilon)=[\epsilon {I}-H_{eq}+ i\sum_{\alpha}W_{\alpha}']^{-1},\label{EqGr1}
\end{equation}
\begin{equation}
\bar{{G}}^{r'}(\epsilon)=[\epsilon {I}-H_{neq}+ i\sum_{\alpha}W_{\alpha}']^{-1}.\label{EqGr2}
\end{equation}
Since $W^{'}_{\alpha}$ is energy independent, we can use following eigen-equations to construct the retarded Green's functions,
\begin{eqnarray}
&&[H_{neq}- i\sum_{\alpha}W_{\alpha}']|\psi_n\rangle=\epsilon_n|\psi_n\rangle\\
&&[H_{neq}+ i\sum_{\alpha}(W_{\alpha}')^{\dagger}]|\phi_n\rangle=\epsilon^*_n|\phi_n\rangle,
\end{eqnarray}
and
\begin{eqnarray}
&&[H_{eq}- i\sum_{\alpha}W_{\alpha}']|\psi^0_n\rangle=\epsilon^0_n|\psi^0_n\rangle\\
&&[H_{eq}+ i\sum_{\alpha}(W_{\alpha}')^{\dagger}]|\phi^0_n\rangle=\epsilon^{0*}_n|\phi^0_n\rangle.
\end{eqnarray}
Then retarded Green's functions can be constructed from their eigenfunctions
\begin{equation}
\tilde{{ G}}^{r'}(\epsilon)=\sum_{n}\frac{|\psi^0_n\rangle\langle\phi^0_n|}{(\epsilon-\epsilon^0_n+i0^+)},\label{EqGr3}
\end{equation}
\begin{equation}
\bar{{ G}}^{r'}(\epsilon)=\sum_{n}\frac{|\psi_n\rangle\langle\phi_n|}{(\epsilon-\epsilon_n+i0^+)}.\label{EqGr4}
\end{equation}

Due to the presence of the time dependent factor $e^{-i(\omega - \epsilon)t}$ in ${A}^{'}_{\alpha}$, the integration in ${A}^{'}_{\alpha}(\epsilon,t)$ can be done analytically by enclosing a contour in the lower half of complex plane,
\begin{equation}
\begin{split}
{A}^{'}_{\alpha}(\epsilon,t)&=\sum_{n}\frac{|\psi_n\rangle\langle\phi_n|}{(\epsilon+\Delta_{\alpha}-\epsilon_n+i0^+)}
+\sum_n \frac{e^{i(\epsilon+\Delta_{\alpha}-\epsilon_n)t}|\psi_n\rangle\langle\phi_n|}{\epsilon-\epsilon_n+i0^+} \times \\
&[\frac{\Delta_{\alpha}}{\epsilon+\Delta_{\alpha}-\epsilon_n+i0^+}-
{\Delta}\sum_{m}\frac{|\psi^0_m\rangle\langle\phi^0_m|}{(\epsilon-\epsilon^0_m+i0^+)}].
\end{split}
\end{equation}

\begin{figure}[!ht]
\centering
\includegraphics[width=8.5cm,height=6cm]{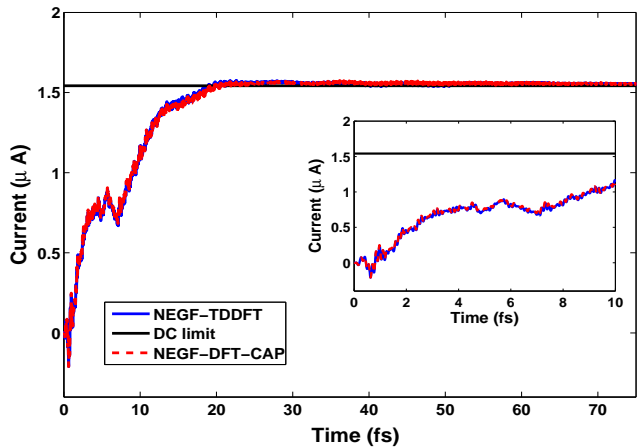}
\caption{The time dependent transient current $I(t)$ versus time with $V=0.0005$ a.u. for one dimensional atomic Al-C$_1$-Al chain. Blue solid line and red dashed line are time transient dependent current calculated by using NEGF-TDDFT method and NEGF-DFT-CAP method; black solid line is the DC current at steady state limit.} \label{fig3}
\end{figure}

It is easy to check that in the initial and asymptotic long time limit ($t\rightarrow \infty$) ${A}^{'}_{\alpha}(t)$ is equal to $\tilde{{G}}^{r'}(\epsilon)$ and $\bar{{G}}^{r'}(\epsilon)$, respectively. After obtaining the ${A}^{'}_{\alpha}(t)$, one can calculate lesser Green's function using Eq. (\ref{Gl}) and hence the time dependent current from Eq. (\ref{tdc}). Since ${A}^{'}_{\alpha}(t)$ is expressed as the summation form at any given time $t$, one only needs to integrate the energy $\omega$ in Eq. (\ref{Gl}) to obtain the time dependent lesser Green's function and hence the transient current $I_{\alpha}(t)$. We can estimate the number of operations in calculating time dependent current $I_{\alpha}(t)$. For a given time, the calculation only involves matrix multiplication as well as the integral over $\omega$ which again can be done using the theorem of residue. Hence the total number of operations is roughly $c N n^3$, an order N algorithm, where $c$ is of order 200 due to the contour integral on the complex plane, $n$ is the dimension of Green's function of the whole system including CAP region, and $N$ is number of time steps.

The major steps for the numerical calculation can be summarized as follows. We first prepare the initial equilibrium and final non-equilibrium self consistent Hamiltonian from a DC calculation. Then we construct the CAP matrix $W_{\alpha}$ with respective to the lead. Once the CAP is constructed, one can compare the transmission coefficients with that obtained by exact method to get an idea how long the CAP region should be. With the good agreement on the transmission coefficient, we can move on to calculate the time dependent current using ${A}^{'}_{\alpha}(t)$.

It is worth mentioning that, in the above discussion, the orthogonal basis set is implicitly assumed to expand the Hamiltonian. So one has to orthogonalize the basis set if non-orthogonal basis such as atomic orbital basis set (LCAO) is used.\cite{Xing1}

\section{Numerical results}
\begin{figure}[!ht]
\centering
\includegraphics[width=7.5cm,height=5.5cm]{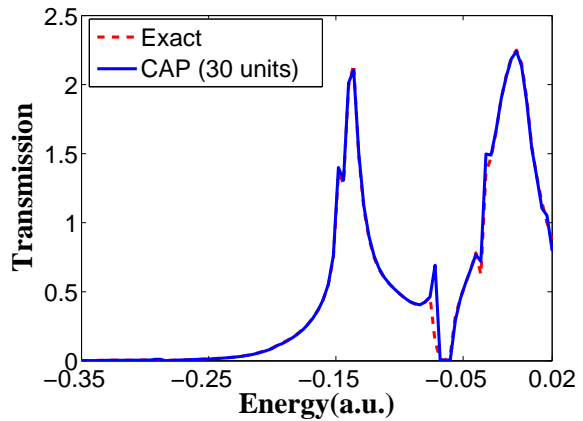}
\caption{The comparison of transmission coefficient of a Carbon chain sandwiched between Al(100) leads. The numerical results calculated by using CAP method with $30$ unit cells in the lead region (blue solid line) is compared with exact method (red dashed line).}\label{fig4}
\end{figure}
In this section, the implementation of our formalism and numerical results of transient current for Al-C$_3$-Al molecular device will be presented. The structure of Al-C$_3$-Al molecular device is shown in figure \ref{fig2}. There are $75$ atoms in the central scattering region and the distance between the Al atom and the nearest carbon atom is equal to $3.78$ a.u. As for the electrodes, there are $9$ aluminum atoms in a unit cell with a finite cross section along ($100$) direction in the semi-infinite aluminum lead.

Our numerical analysis is based on the state-of-the-art first principles quantum transport package MATDCAL.\cite{Waldron1,Waldron2} Specifically, a linear combination of atomic orbitals (LCAO) is employed to solve KS equations. The exchange-correlation is treated at the LDA level and the nonlocal norm-conserving pseudopotential\cite{Troullier} is used to define the atomic core. The density matrix is constructed in orbital space and the effective potential is obtained in real space by solving the Poisson equation. The accuracy in the self-consistent iteration is numerically converged to $10^{-4}$ eV. The initial equilibrium and final non-equilibrium Hamiltonians are prepared using MATDCAL package.

In the following, the case of upward step-like pulse ($V_L(t)=-V_R(t)=\theta(t)V$) applied on both leads will be considered. In order to satisfy the current conservation condition, we will plot the time dependent current in terms of $I(t)=[I_L(t)-I_R(t)]/2$\cite{Xing1}.

Before presenting our numerical results, we have calculated the transient current using two different approaches to test the accuracy of our present scheme. One is based on the NEGF-TDDFT method proposed in Ref. \cite{lei} which is an order $N^2 (log_2 N)^2$ algorithm and other one is our proposed formalism in this paper termed as NEGF-DFT-CAP. Here we take one dimensional Al-C$_1$-Al atomic chain (where both leads are one-dimensional Al chain) as a toy molecular device and apply a step-like pulse to test numerical implementation of our formalism. As shown in figure \ref{fig3}, transient current calculated from two different methods agree well with each other. In addition, the transient current approaches the DC steady state value obtained by using Landauer-B\"{u}ttiker formula in the long time limit. The insert figure shows the early time behavior of the transient current.

\begin{figure}[!ht]
\centering
\includegraphics[width=8.2cm,height=5.8cm]{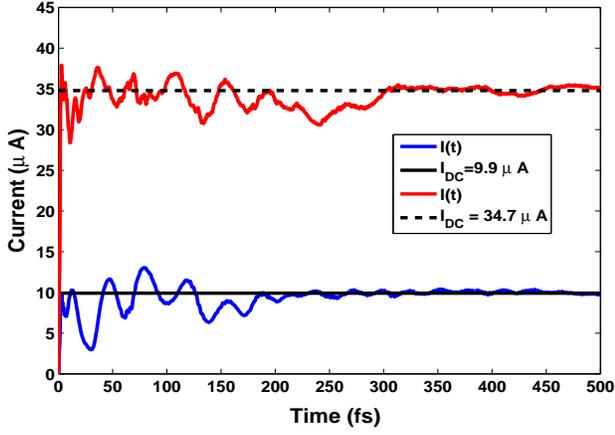}
\caption{The time dependent transient current $I(t)$ versus time with different bias voltages for Al-C$_3$-Al molecular device. The blue and black solid lines correspond with time dependent transient current and DC current at steady state for $V=0.0025$ a.u., respectively. The red solid and black dashed lines are time dependent transient current and DC current at steady state for $V=0.01$ a.u., respectively.}\label{fig5}
\end{figure}

Now let us study the real molecular device Al-C$_3$-Al. First of all, we have to compare the transmission coefficient by using CAP method and exact method to make sure that the CAP potential is added correctly. As you can see in figure \ref{fig4}, the CAP results agrees well with that calculated by exact method.

After comparing the accuracy of transmission coefficient of CAP method, we are ready to study the transient current of Al-C$_3$-Al device. We calculate the transient current under two different bias voltages. The numerical results are plotted in figure \ref{fig5}. We have serval observations: (1) the switch-on time is roughly 2 fs; (2) the relaxation time is roughly $210$ fs for $V=0.0025$a.u. and $320$ fs for $V=0.01$a.u.; (3) the transient current is on the same order of magnitude as that of the DC steady state limit. In the early time, there are some irregular oscillations in the transient current. At the long time limit, the transient current approaches to the correct DC limit. Moreover, more oscillations occur with the increase of bias voltage. The nature of the oscillation can be attributed to the resonant states of the system\cite{Bin1}.

\section{Conclusion}

To summarize, we have proposed an order N first-principle formalism to study dynamical response of molecular device due to the time dependent step like external bias. Our formalism is based on NEGF combined with DFT as well CAP method. The use of CAP allows us to calculate transient current efficiently. Comparing with the previous NEGF-TDDFT schemes, the computational cost of current scheme is much less. Detailed computational procedures for first principles transient current calculation were discussed which is very easy to implement. As an illustration, we have calculated the transient current of Al-C$_3$-Al molecular devices from first principles.

\bigskip

\noindent{$^{*)}$ Electronic address: jianwang@hku.hk}

\section{Acknowledgments}

We gratefully acknowledge the support from Research Grant Council (HKU 705611P) and University Grant Council (Contract No. AoE/P-04/08) of the Government of HKSAR. This research is conducted using the HKU Computer Centre research computing facilities that are supported in part by the Hong Kong UGC Special Equipment Grant (SEG HKU09).

\appendix
\section{Derivations for $G^{r'}_{LR}$}
According to the definition of retarded Green's function,
\begin{align}
\left[\begin{array}{ccc}
(g^{r'}_{LL})^{-1} & -H_{LC} & 0\\
-H_{CL} & E-H_{CC} & -H_{CR}\\
0 & -H_{RC} & (g^{r'}_{RR})^{-1}\\
\end{array}\right]
\left[\begin{array}{ccc}
G^{r'}_{LL} & G^{r'}_{LC} & G^{r'}_{LR}\\
G^{r'}_{CL} & G^{r'}_{CC} & G^{r'}_{CR}\\
G^{r'}_{RL} & G^{r'}_{RC} & G^{r'}_{RR}\\
\end{array}\right]
=\left[\begin{array}{ccc}
1_{L} & 0 & 0\\
0 & 1_{C} & 0\\
0 & 0 & 1_{R}\\
\end{array}\right],\label{Appendix1}
\end{align}
we have
\begin{equation}
G^{r'}_{LR}=g^{r'}_{LL}H_{LC}G^{r'}_{CR}.
\end{equation}
To find $G^{r'}_{CR}$ we note that the advanced Green's function can be obtained by replacing superscript $r$ into $a$ in equation (\ref{Appendix1}). We have
\begin{equation}
\begin{split}
&G^{a'}_{RC}=g^{a'}_{RR}H_{RC}G^{a'}_{CC},\\
&G^{r'}_{CR}=G^{r'}_{CC}H_{CR}g^{r'}_{RR},\label{Appendix2}
\end{split}
\end{equation}
where we have used the fact that $G^{r'}_{CR}=(G^{a'}_{RC})^\dagger$. Finally we combine equations (\ref{Appendix1}) and (\ref{Appendix2}) to arrive at
\begin{equation}
G^{r'}_{LR}=g^{r'}_{LL}H_{LC}G^{r'}_{CC}H_{CR}g^{r'}_{RR}.
\end{equation}

\section{Derivations for $A'_{\alpha_{CC}}$}
In this appendix, we will derive the expression of $A'_{\alpha_{CC}}$ in central region within CAP method to confirm $A'_{\alpha}$ given in equation (\ref{EqA}). Starting from equation (\ref{EqA}), we have
\begin{equation}
\begin{split}
&{A}^{'}_{\alpha_{CC}}(\epsilon,t)=\bar{{G}}^{r'}_{CC}(\epsilon+\Delta_{\alpha})
-\int\frac{d\omega}{2\pi i}\frac{e^{-i(\omega-\epsilon)t}}{\omega-\epsilon+\Delta_{\alpha}-i0^+}\\&
[\bar{{G}}^{r'}_{CC}(\omega+\Delta_{\alpha})\frac{\Delta_{\alpha}}{\omega-\epsilon-i0^+}+BB],\label{AppendixEqA}
\end{split}
\end{equation}
with
\begin{equation}
\begin{split}
BB\equiv\sum_{\beta=L,C,R}\bar{{G}}^{r'}_{C\beta}(\omega+\Delta_{\alpha})
{\Delta_{\beta \beta}}\tilde{{G}}^{r'}_{\beta C}(\epsilon)
\end{split}
\end{equation}
where $\Delta_{CC}$ is the internal potential change in the central region due to the external bias in the leads and $\Delta_{\beta\beta}=\Delta_{\beta}{1}_{\beta}$ with $\beta=L,R$ is the bias applied in lead $\beta$. Furthermore, $BB$ can be separated into two parts,
\begin{equation}
\begin{split}
BB&=\bar{{G}}^{r'}_{CC}{\Delta_{CC}}\tilde{{G}}^{r'}_{C C}+\sum_{\beta=L,R}\Delta_{\beta}\bar{{G}}^{r'}_{C\beta}
\tilde{{G}}^{r'}_{\beta C}\\
&=\bar{{G}}^{r'}_{CC}{\Delta_{CC}}\tilde{{G}}^{r'}_{C C}+\sum_{\beta=L,R}\Delta_{\beta}\bar{{G}}^{r'}_{CC}H_{C\beta}{\bar g}^{r'}_{\beta\beta}
{\tilde g}^{r'}_{\beta\beta}H_{\beta C}\tilde{{G}}^{r'}_{CC},
\end{split}
\end{equation}
where we have used $\bar{{G}}^{r'}_{C\beta}=\bar{{G}}^{r'}_{CC}H_{C\beta}\bar{g}^{r'}_{\beta\beta}$ and $\tilde{{G}}^{r'}_{\beta C}=\tilde{g}^{r'}_{\beta\beta}H_{\beta C}\tilde{{G}}^{r'}_{CC}$. According to the retarded Green's function of lead $\alpha$ in equation (\ref{EqLeadGr}), we have
\begin{equation}
\begin{split}
{\bar g}^{r'}_{\beta\beta}{\tilde g}^{r'}_{\beta\beta}&=\frac{1}{\omega+\Delta_{\alpha}-\Delta_{\beta}-H'_{\beta\beta}}
\frac{1}{\epsilon-H'_{\beta\beta}}\\
&=\frac{1}{\epsilon-\omega-\Delta_{\alpha}+\Delta_{\beta}}[\frac{1}{\omega+\Delta_{\alpha}-
\Delta_{\beta}-H'_{\beta\beta}}-\frac{1}{\epsilon-H'_{\beta\beta}}].
\end{split}
\end{equation}
Therefore, $BB$ becomes
\begin{equation}
\begin{split}
BB=\bar{{G}}^{r'}_{CC}{\Delta_{CC}}\tilde{{G}}^{r'}_{C C}-\sum_{\beta=L,R}\Delta_{\beta}\bar{{G}}^{r'}_{CC}\tilde{\Upsilon}^R_{\alpha\beta}(\epsilon,\omega)
\tilde{{G}}^{r'}_{CC},\label{AppendixEqB}
\end{split}
\end{equation}
where we have defined
\begin{equation}
\begin{split}
\tilde{\Upsilon}^R_{\alpha\beta}(\epsilon,\omega)\equiv\frac{\tilde{\Sigma}^{R'}_{\beta}(\epsilon)
-\tilde{\Sigma}^{R'}_{\beta}(\omega+\Delta_{\alpha}-\Delta_{\beta})}
{\epsilon-\omega-\Delta_{\alpha}+\Delta_{\beta}}.
\end{split}
\end{equation}
Finally, by plugging equation (\ref{AppendixEqB}) for $BB$ into equation (\ref{AppendixEqA}), one can easily find that the final expression for ${A}^{'}_{\alpha_{CC}}$ is the same as the expression for $A_{\alpha_{CC}}$ given in Ref. \cite{Maciejko}.


\section{References}

\begin{thebibliography}{10}

\bibitem{Aviram}
Aviram A and Ratner M 1974 Chem. Phys. Lett. {\bf 29} 277

\bibitem{Chen}
Chen J, Reed M A, Rawlett A M and Tour J M 1999 Science {\bf 286} 1550

\bibitem{Collier}
Collier C P \emph{et al} 1999 Science {\bf 285} 391

\bibitem{Joachim}
Joachim C, Gimzewski J K and Aviram A 2000 Nature {\bf 408} 541

\bibitem{Heath}
Heath J R and Ratner M A 2003 Phys. Today {\bf 56} 43

\bibitem{Petta}
Petta J R, Slater S K and Ralph D C 2004 Phys. Rev. Lett. {\bf 93} 136601

\bibitem{Tao}
Tao N J Nature Nanotechnology 2006 {\bf 1} 173

\bibitem{Ventra}
Ventra M D, Pantelides S T and Lang N D 2000 Phys. Rev. Lett. {\bf 84} 979

\bibitem{Taylor}
Taylor J, Guo H and Wang J 2001 Phys. Rev. B {\bf 63} 245407; {\it ibid.} {\bf 63}  121104

\bibitem{Brandbyge}
Brandbyge M, Mozos J L, Ordejon P, Taylor J and Stokbro K 2002 Phys. Rev. B {\bf 65} 165401

\bibitem{xue}
Xue Y, Datta S and Ratner M A 2002 Chem. Phys. {\bf 281} 151

\bibitem{Kaun1}
Kaun C C, Larade B and Guo H 2003 Phys. Rev. B {\bf 67} 121411R

\bibitem{Kaun2}
Kaun C C and Guo H 2003 Nano Lett. {\bf 3} 1521

\bibitem{Frederiksen}
Frederiksen T, Brandbyge M, Lorente N and Jauho A P 2004 Phys. Rev. Lett. {\bf 93} 256601

\bibitem{Lee}
Lee T, Wang W, Klemic J F, Zhang J J, Su J and Reed M A 2004 J. Phys. Chem. B {\bf 108} 8742

\bibitem{Jauho1}
Wingreen N S, Jauho A P and Meir Y, 1993 Phys. Rev. B {\bf 48}, 8487

\bibitem{Stefanucci}
Stefanucci G, Kurth S, Rubio A, and Gross E K U 2008 Phys. Rev. B {\bf 77} 075339

\bibitem{lei}
Zhang L, Xing Y X and Wang J 2012 Phys. Rev. B {\bf 86} 155438

\bibitem{Gross1}
E. Runge and E. K. U. Gross, Phys Rev Lett {\bf 52} 997 (1984).

\bibitem{Maciejko}
Maciejko J, Wang J and Guo H 2006 Phys. Rev. B {\bf 74} 085324

\bibitem{Zheng}
Zheng X, Wang F, Yam C Y, Mo Y and Chen G H 2007 Phys. Rev. B {\bf 75} 195127

\bibitem{Bin1}
Wang B, Xing Y X, Zhang L and Wang J Phys. Rev. B {\bf 81} 2010 121103(R)

\bibitem{Kosloff}
Kosloff R and Kosloff D 1986 J. Comput. Phys. {\bf 63} 363

\bibitem{Henderson}
Henderson T M et al 2006 J. Chem. Phys. {\bf 125} 244104

\bibitem{Zhang}
Zhang X G et al 2007 Phys. Rev. B {\bf 76} 035108

\bibitem{Varga1}
Driscoll J A and Varga K 2008 Phys. Rev. B {\bf 78} 245118

\bibitem{Varga2}
Varga K 2011 Phys. Rev. B {\bf 83} 195130

\bibitem{Sancho}
Lopez-Sancho et al 1984 J. Phys. F: Met. Phys. {\bf 14} 1205\\
Lopez-Sancho et al 1985 J. Phys. F: Met. Phys. {\bf 15} 851

\bibitem{Sanvito}
Sanvito S et al 1999 Phys. Rev. B {\bf 59} 11936

\bibitem{Manolopoulos}
Gonzalez-Lezena T, Rackham E J and Manolopoulos D E 2004 J. Chem. Phys. {\bf 120} 2247

\bibitem{datta}
Datta S 1995 Electronic Transport in Mesoscopic Systems (Cambridge University Press)

\bibitem{Varga3}
Cook G B, Dignard P and Varga K 2011 Phys. Rev. B {\bf 83} 205105

\bibitem{Xing1}
Xing Y X, Wang B and Wang J 2010 Phys. Rev. B {\bf 82} 205112

\bibitem{Jauhobook}
Huang H and Jauho A -P, 1996 Quantum Kinetics in Transport and Optics of Semi-conductors (Springer-Verlag Press)

\bibitem{Jauho}
Jauho A -P, Wingreen N S and Meir Y 1994 Phys. Rev. B {\bf 50} 5528

\bibitem{Waldron1}
Waldron D, Haney P, Larade B, MacDonald A and Guo H 2006 Phys. Rev. Lett. {\bf 96} 166804

\bibitem{Waldron2}
Waldron D, Timoshevskii V, Hu Y, Xia K and Guo H 2006 Phys. Rev. Lett. {\bf 97} 226802

\bibitem{Troullier}
Troullier N and Martins J L 1991 Phys. Rev. B {\bf 43} 1993


\end{thebibliography}
\end{document}